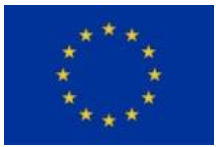
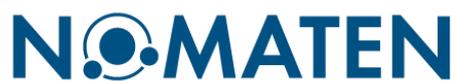

This work was carried out in whole or in part within the framework of the NOMATEN Centre of Excellence, supported from the European Union Horizon 2020 research and innovation program (Grant Agreement No. 857470) and from the European Regional Development Fund via the Foundation for Polish Science International Research Agenda PLUS program (Grant No. MAB PLUS/2018/8), and the Ministry of Science and Higher Education's initiative "Support for the Activities of Centers of Excellence Established in Poland under the Horizon 2020 Program" (agreement no. MEiN/2023/DIR/3795).

The version of record of this article, first published in Separation and Purification Technology, Volume 351, 24 December 2024, 128074, is available online at Publisher's website: https://doi.org/10.1016/j.seppur.2024.128074

# Evaluation of tetracycline photocatalytic degradation using NiFe2O4/CeO2/GO nanocomposite for environmental remediation: In silico molecular docking, Antibacterial performance, degradation pathways, and DFT calculations

Misbah latif [1], Raziq Nawaz[2, 3], Muhammad Hammad Aziz[1, 2,*], Muhammad Asif[1], Fatima Noor[1], Amil Aligayev[2, 3, 4], Syed Mansoor Ali[5], Manawwer Alam[6], Stefanos Papanikolaou[4], Qing Huang[2, 3,**]

[1]Department of Physics, COMSATS University Islamabad, Lahore Campus, 54000, Lahore, Pakistan

[2]Institute of Intelligent Machines, Hefei Institutes of Physical Science, Chinese Academy of Sciences, Hefei 230031, China

[3]University of Science and Technology of China, Hefei, 230026, China

[4]NOMATEN Centre of Excellence, National Centre for Nuclear Research, 05-400 Swierk/Otwock, Poland

[5]Department of Physics and Astronomy, College of Science, P.O. BOX 2455, King Saud University, Riyadh 11451, Saudi Arabia

[6]Department of Chemistry, College of Science, P.O. BOX 2455, King Saud University, Riyadh 11451, Saudi Arabia

Corresponding Authors:

*Muhammad Hammad Aziz; hammadaziz@cuilahore.edu.pk

**Qing Huang: huangq@ipp.ac.cn

## Abstract

Graphene-based nanostructures with distinct structural and physicochemical characteristics may be able to photodegrade antibiotics effectively. Herein, this study reports the successful synthesis of NiFe2O4/CeO2/GO nanocomposite (NC) by anchoring NiFe2O4/CeO2 to the surface of GO (Graphene oxide). All state-of-the-art characterization techniques investigated the nanostructure, crystallinity, phonon modes, chemical composition analysis, elemental composition, surface area, magnetic properties, and optical band gap. Hydrothermal approach assisted NiFe2O4/CeO2/GO catalyst showed better charge carrier separation and prompted the tetracycline (TC-HCl) photocatalytic degradation under visible light. Following 90 minutes of exposure to visible light, NiFe2O4/CeO2/GO nanocomposite demonstrated superior photocatalytic activity, with a TC-HCl degradation rate of 95%. Reasonable mechanisms of tetracycline degradation were proposed where the •OH and •$O^-$ played a leading role based on identified intermediates. Moreover, tetracycline photodegradation intermediates and the optimal pathway were identified using LC-MS spectrometry. This study also performed Density Functional Theory (DFT) calculations for the prepared materials to validate the experimental data. In vitro, antibacterial studies were consistent with the molecular docking investigations of the NiFe2O4/CeO2/GO nanocomposite against DNA gyrase and FabI from Escherichia coli (E. coli) and Staphylococcus aureus (S.

aureus). Lastly, the outcomes revealed a new potential for NiFe2O4/CeO2/GO nanocomposite for improved photocatalytic performance, making it a promising photocatalyst for wastewater treatment.

**Keywords:**
Graphene Oxide; Tetracycline Hydrochloride (TC-HCl); Photocatalyst; Density Functional Theory (DFT); Photocatalysis; Antibacterial activity.

Introduction

Pharmaceuticals in aquatic ecosystems are mostly sourced via the constant discharge of wastewater from treatment plants, with additional sources including waste from agricultural areas, aquaculture activities, and cattle ranches [1, 2]. Pharmaceuticals are synthetic chemicals developed for clinical application as therapeutic medications. Because of how they are constructed, medicines can cause significant physiological changes in living organisms even when administered at very low doses. These substances, whether excreted as metabolites or unaltered, persist for long periods in the atmosphere as they are not physiologically destroyed or reduced [3, 4].

Antibiotic resistance genes (ARG) have been amassing as antibiotic usage has skyrocketed over the past few decades, posing a hazard to animal, plant, and human ecosystems [5-7]. The tetracycline antibiotic family is the second most extensively used class of pharmaceuticals globally [8, 9]. It is employed in various areas, including medicine, aquaculture, and animal husbandry. Tetracycline hydrochloride is often utilized to cure and avoid bacterial infections in rats, raptors, and amphibians [10]. Since urbanization and industry have expanded rapidly, there has been an overuse of TC-HCl antibiotics and poor management, which has led to more serious water pollution and poses a serious hazard to both humans and the environment. Due to its mutagenic and teratogenic effects, it has garnered much attention [11, 12]. In order to address the environmental problems associated with tetracycline, scientists have developed and implemented a number of approaches, including photocatalysis, adsorption, biological removal, and others. This has improved the water quality and capacity for self-purification that tetracycline has enabled. [13, 35 14]. Furthermore, magnetic ferrites-based photocatalyst has become an emerging candidate to decompose antibiotics in wastewater [15-17] due to its low cost, lack of secondary pollutants, and environment-friendly approach. As a result, the main goal is to develop a photocatalyst with excellent degrading performance for TC with efficient photo-generated carrier separation [18-21]. Significant interest has been shown in heterogeneous photocatalytic oxidation as a potentially effective technique for decomposing refractory environmental contaminants. Doping with transition metal ions [22] and combining them with other narrow-band gap semiconductors functioning as photocatalysts [23] have been the subjects of several recent investigations. Because of their suitable band gap, strong photocatalytic activity, and chemical stability, several additional materials, including $CeO_2$, $SnO_2$, $ZrO_2$, $Sm_2O_3$, $Fe_2O_3$, $TiO_2$, $V_2O_5$, $WO_3$, $ZnO$, $CdS$, $Bi_2O_3$, $Sb_2O_3$, $Al_2O_3$, and $MoO_3$, have been employed as conventional photocatalysts [24]. According to previous studies, a significant number of binary composites, like $V_2O_5$-$TiO_2$, $V_2O_5$-$ZnO$, $Au/Fe_2O_3$, and $BiVO_4/FeVO_4$, and an insignificant amount of ternary nanocomposites, like $V_2O_5$-$WO_3/TiO_2$, $Ni/MgO$-$ZrO_2$, and $N$-$Fe_2O_3/FeVO_4$ [25, 26], have been developed and synthesized for organic pollution under observations. Phenols are among the ecological contaminants that are largely excluded by semiconductor particles containing metal. Numerous studies indicate that cations such as $Fe^{3+}$, V, Ni, Pt, $Cr^{3+}$, $Cu^{2+}$, Co, Au, Pd, or other ions might boost the photoactivity of nanoparticles [27].

Magnetic ferrites such as spinel ferrites have a structural formula MFe2O4 (M =Ag, Au, Mg, Mn, Co, Ni, Cu, and Ce). Magnetic ferrites are utilized in different applications including catalysis, drug delivery, magnetic resonance imaging (MRI), microwave devices, photocatalysis, and water treatment. Their structures and characteristics are mainly determined by the arrangement of the cations in the tetrahedral and octahedral displays. Due to their high chemical stability, magnetic ferrites have recently attracted the interest of numerous scientists. Currently, researchers have been interested in magnetic ferrite nanocomposites because of their exceptional chemical stability, excellent reusability, and also magnetically detached. Substantial research has been put into enhancing the structure of magnetic ferrite nanocomposite to address the key issues that prevent their effective use in wastewater treatment. Numerous studies have also demonstrated magnetic ferrite's potential for use in photocatalytic processes that are driven by visible light. Nickel ferrite is the most significant magnetic material among various ferrites due to its thermal stability, electromagnetic performance, chemical stability, and high Curie temperature [28-30], The photocatalytic activity of nickel ferrite can be increased by incorporating the metal oxides nanoparticles such as ZnO, $CeO_2$, $MnO_2$, NiO, MgO, $SnO_2$, CuO, and $ZrO_2$, etc, which facilitate more effective separation of the photo-generated charge carriers when exposed to visible light [31]. Especially, Cerium oxide has received a lot of interest because of its wide bandgap (Eg = 3.19 eV). Also, $CeO_2$ is a rare earth dioxide that is also used as a heterogeneous photo-catalyst and contains redox sequences between the $Ce^{3+}$ (reduced) and $Ce^{4+}$ (oxidized) oxidation states [32-34].

Graphene is considered a two-dimensional carbon network with sp2 hybridization and is recognized due to its outstanding electrical properties, effective electron mobility, high surface area, and chemical stability [35]. Graphene's exceptional qualities can be leveraged by combining it with other nanomaterials to boost catalytic activities. By accepting photo-generated electrons and minimizing charge carrier recombination, graphene has the potential for very efficient catalytic destruction of contaminants. Graphene-related nanomaterials are also utilized to increase the catalytic activity because they have a higher absorption range and enhanced charge separation and transport [36]. Graphene-based nanocomposite acts as a better photocatalyst as reduces electron-hole recombination which boosts the electron transfer rate, and improves the adsorption of chemical species through π-π interactions. Liang et al. demonstrated the visible-light-active $NiFe_2O_4$-rGO composite photocatalyst, which exhibited remarkable photocatalytic performance [37]. Deng et al. revealed that $rGO/SnS_2/ZnFe_2O_4$ enhanced the photocatalytic effectiveness by the presence of rGO [38]. The oxygenated moieties in GO create defects and functional sites, thereby influencing the material's electronic structure. Its high surface area provides an ideal platform for the interaction and stabilization of the composite components, ultimately influencing the electronic and catalytic properties of the $NiFe_2O_4/CeO_2/GO$ system. To systematically elucidate these impacts, Density Functional Theory (DFT) computations were also performed in this study.

In this view, Ultimate research in the realm of photocatalysis technology includes developing and investigating high-performance materials that respond to sunlight. The widespread co- existence of harmful medicines and metal ions in wastewater makes the development of a multifunctional photocatalyst imperative for the high-efficiency treatment of these pollutants. Ion intercalation and functionalization of the surface have been widely used to construct catalysts that include oxyanion modification into metal ferrites. The majority of oxyanions, such as $CO_2-$ and $NO-$, may be intercalated directly in the metal ferrites catalysts' interlayer gaps. Furthermore, the thorough use of photocatalysis is severely limited by the quick pairing of photoelectron-hole pairs [39-41]. As a result, to solve this constraint, researchers have used a variety of approaches, such as morphological control, semiconductor composite, precious metal surface modification, and other ways, to change the metal oxides with another composite material [42-45]. Among them, it has been shown that the heterojunction and three-dimensional hierarchical frameworks greatly increase the utilization of light absorption and encourage carrier separation, which permits quick interfacial charge transfer and substantially upgrades the photoactivity. It is noteworthy that the majority of pertinent research focuses on organic pollutants, although in practice, wastewater treatment involves both organic materials and microbiological pathogens. Consequently, further thorough study is required to create a novel multifunctional degradation and disinfection catalyst based on ferrite nanocomposites that are linked to graphene [46,47].

Therefore, In this study, $NiFe_2O_4/CeO_2/GO$ nanocomposite was used to explore the photocatalytic performance of tetracycline. The findings demonstrated that $NiFe_2O_4/CeO_2/GO$ has greater stability and catalytic activity than $NiFe_2O_4/CeO_2$. Besides, the physical characteristics and physiochemical properties of $NiFe_2O_4/CeO_2/GO$ are characterized by state of art techniques. This investigation showed that the combination of $NiFe_2O_4/CeO_2$ and GO produced excellent photocatalytic activity, with rapid separation and quick transfer of photo-generated carriers. Photodegradation of TC-HCl into smaller molecular intermediates was investigated via LC-MS. In silico molecular docking simulations were conducted to investigate the potential bactericidal of the $NiFe_2O_4/CeO_2/GO$ nanocomposite against DNA gyrase and FabI from E. coli and S. aureus. Overall, this study offers an effective approach by using $NiFe_2O_4/CeO_2/GO$ as an excellent photocatalyst for wastewater treatment and industrial purposes as well.

## 2. Experimental procedure

### 2.1 Chemicals

Deionized (D.I) water, Tetracycline (TC-HCl), Nickel chloride Hexahydrate (NiCl2.6H2O), Sodium Hydroxide (NaOH), Iron chloride Hexahydrate (FeCl2.6H2O), Graphite, Cerium Nitrate (Ce(NO3)3.6H2O) were used in this research.

### 2.2 Synthesis of $NiFe_2O_4$, $NiFe_2O_4/CeO_2$ and $NiFe_2O_4/CeO_2/GO$ via hydrothermal approach

Nickel chloride Hexahydrate (3 g) and Iron chloride (6g) were dissolved in 500 mL DI water and agitated for 120 minutes. Afterward, NaOH (2g) was dissolved in 25 mL and further added drop by drop while constantly agitating the mixture. The resultant mixture was then subjected to heat at 180 °C for 6 hrs. in a stainless steel autoclave. After being separated, the solution was gently rinsed with DI water 5 times and then calcinated at 400°C for 6 hrs. Next, $NiFe_2O_4/CeO_2$ nanocomposite was synthesized via the hydrothermal method. Precursors such as 2 grams of (NiCl2.6H2O), 4.5 grams of ($FeCl_2.6H_2O$), and 2.5 grams of ($Ce(NO_3)_3.6H_2O$) 0were mixed in D.I (750 mL) and agitated for 1 hr. The obtained mixture was stirred for a further 30 minutes while NaOH was mixed drop by drop again. After that, the obtained solution was then placed into a Teflon-lined autoclave to go through the hydrothermal process at 200°C for 6 hrs. Later, the obtained powder was separated and thoroughly washed with D.I. water at least 5 times before calcinating for 12 hours at 500°C. GO was prepared by modifying Hummer's process [48]. Initially, 200 mg of graphene oxide was mixed in distilled water of 50 mL. Further, the GO solution was mixed with 100 mL of NiFe$_2$O$_4$/CeO$_2$ (1g) composite solution.

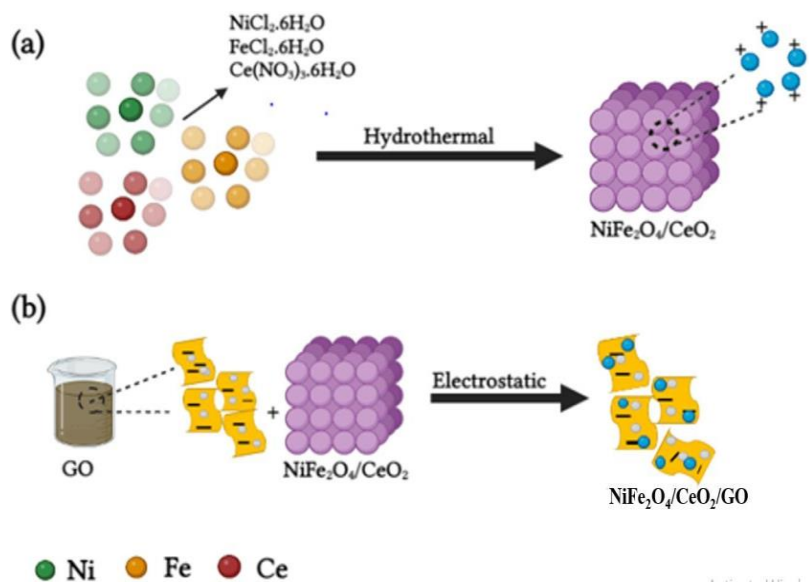

Figure 1: Schematic representation of NiFe2O4/CeO2 and NiFe2O4/CeO2/GO nanocomposite via hydrothermal process

The suspension was subsequently put through a 1-hour ultrasonication treatment at 60°C. Now, the solution was continuously stirred for 1 hour and further kept for 12 hours for aging. For hydrothermal treatment, the resulting solution was then subjected to a stainless steel autoclave at 180°C for 12 hrs. Lastly, the nanocomposite solution was centrifuged and calcinated at 400 °C for 4 hours to obtain the NiFe2O4/CeO2/GO nanocomposite as illustrated in Figure 1.

**2.3 Photocatalytic measurements**

The photocatalytic activity was carried out at room temperature (25 °C) in quartz glass reactors. A 300 W Xe lamp with a 420 nm wavelength light source was used, and the surface was illuminated at a 50 W/m2 intensity. Typically, photocatalyst (50 mg) was assorted to 100 mL of reaction fluid at 20 mg/L of tetracycline concentration in a reactor. The reaction mixture was agitated in the dark for about 30 minutes to achieve adsorption-desorption equilibrium. The suspension was filtered using 0.45 μm filters after being removed at regular intervals of 1 mL during irradiation. TC-HCl concentration was obtained by UV-Vis spectrophotometer in the collected solution while TC-HCl intermediates were evaluated using an LC-MS. A 250 mL mixture of DI water and 100 milligrams of NiFe2O4/CeO2/GO catalyst was let to stand in the dark for half an hour. The resulting solution (3 mL) was taken at intervals of ten minutes, and they were stirred while being exposed to visible light. Under the specific protocols (pH level= 5, TC-HCl (20 mg/L), catalyst dose (1.5 g/L) & light exposure for 90 min). Nanocomposite catalytic properties were evaluated using pH (3.0, 5.0, 7.0, and 9.0) and co-existing anions ($H_2PO^-$, $Cl^-$, $HCO^{2-}$, and 4 3 $NO-$) and optimized under simulated visible light irradiation. IPA (2.0 mmol) for hydroxyl radicals, silver nitrate (2.0 mmol) for electrons, AO (2.0 mmol) for holes, and BQ (0.5 mmol) for superoxide anions were used as scavenging chemicals for trapping reactive radicals.

## 2.4 Recycling Experiments:

$NiFe_2O_4/CeO_2/GO$ photocatalysts were recovered after the reaction was finished by filtering through polypropylene membranes and washing with acetone for five cycles while sonicated for five minutes. Following that, 5 mL of acetone was poured into the entire filtering membrane with the black cake on top in a beaker to produce a stable suspension that could be securely transferred to the storage vial. The sample was then dried using a vacuum. Following that, all of the chemicals and the solvent were added to the heterogeneous catalysts before starting a fresh catalytic cycle.

## 2.5 Antibacterial Activity:

NiFe2O4/CeO2/GO antibacterial activity was evaluated against the bacterial strains that are commonly found in wastewater streams. The nanocomposite's bactericidal potential was investigated using the in-vitro disc diffusion method. The antibacterial efficacy of NiFe2O4/CeO2/GO was examined against gram-positive Staphylococcus aureus (S. aureus) and gram-negative Escherichia coli (E. coli) bacterial strains. To prepare the glass petri plate for analysis, 20 mL of Muller Hinton agar coating was applied. Once the agar solidified, the designated type's 24-hour-old bacterial culture was swabbed onto the agar surface. NiFe2O4/CeO2/GO combined circular discs (100 μg/disc) soaked in 10, 20, 30, and 40 μg/mL of the NiFe2O4/CeO2/GO NC solution were facially added to this plate. chloramphenicol (3 μg/disc) was used as a positive control. The compounds were allowed to diffuse in the plates for 30 minutes, after which they were conditioned for 24 hours at 37 °C in an incubator. To determine the antibacterial activity of the NCs and controls, the clear zone of inhibition surrounding each disc was then assessed.

## 2.6 Molecular Docking Analysis

The molecular docking estimations of the $NiFe_2O_4/CeO_2/GO$ nanocomposite focused on enzymes involved in nucleic acid and fatty acid biosynthesis pathways, which are considered promising targets for antibiotic development [49]. The AutoDock Vina software was utilized to perform molecular docking [50]. The 3D structural coordinates of the enzyme targets were obtained from the Protein Data Bank (PDB) with the following identification numbers: 5MMN for DNA gyrase [51], 1MFP [52], for enoyl-[acyl carrier protein] reductase (FabI) from E. coli, and 6TBC [53], for enoyl-[acyl carrier protein] reductase (FabI) from S. aureus. The analysis of the docked complexes and the 3D representation of binding patterns between the nanocomposite and active site residues were carried out using Discovery Studio, AutoDock Vina Visualizer, and PyMOL in conjunction [54]. Consistent with our earlier studies, 3D structures were analyzed using ChemSketch or SYBYL-X 2.0 to assess binding affinities with the active residues of particular proteins [55,56].

## 2.7 Characterizations

X-ray diffraction (XRD) was used to obtain the crystalline size via PANalytical X'Pert-PRO. Microstructures were investigated on FE-SEM (Hitachi; S-4500, Japan) while element analysis was explored via energy-dispersive X-ray spectroscopy (EDX) utilizing a 200 kV accelerating voltage. XPS analysis was determined on a photoelectron spectrometer (ESCALAB 250Xi). The chemical interaction of prepared materials was revealed by Fourier transform infrared (FTIR) spectroscopy. The Raman analysis was measured by a Raman spectrometer (Xplora Plus, France) to determine the Raman modes. After degassing the samples for 14 hours at 130 °C, the BET porosity was determined using the BELSORP MAX II instrument. The optical band gaps and absorption ranges of the prepared materials were attained via UV-Vis spectroscopy (Shimadzu, UV-2450, Japan). A vibrating sample magnetometer (Model-7410; Lake Shore; USA) was employed for the evaluation of the magnetization data. HPLC mass spectrometer (Agilent 1260 series, UK) was utilized to determine the TC-HCl intermediates produced during the photocatalytic activity. Electron spin resonance (ESR) was done on a spectrometer (Bruker ER200-SRC, Germany) to evaluate free radicals (•OH, $•O^-$).

## 3. Results and discussions

### 3.1 Characterization Techniques and Analysis

In Figure 2, XRD was used with a Cu-Kα radiation source (λ = 0.15402 nm) to investigate the crystalline structure of NiFe2O4, NiFe2O4/CeO2, and NiFe2O4/CeO2/GO. XRD pattern for CeO2 has the following distinctive planes (111), (200), (220), (311), (222), (400), (331), and (420) corresponds to 28.74°, 33.2°, 47.4°, 56.3°, 59.4°, 70.8°, 75.2° and 79.3° which recognized by a cubic fluorite structure as correspond to JCPDS No. 34-0394 and belongs to the space group Fm3m in Figure S1 [57, 58]. Moreover, Crystal planes of NiFe2O4 (111), (220), (311), (400), (422), (511), and (440) belong to 18.7°, 33°, 35.3°, 43.3°, 53.7°, 57.4°, 61.9° are interpreted to signify the cubic structure of NiFe2O4 as matched with JCPDS No. 86-2267 [59, 60]. Furthermore, XRD investigation of NiFe2O4/CeO2 revealed the crystallinity as perceived in Figure 2. In NiFe2O4/CeO2/GO nanocomposite, the diffraction peak of GO reveals at 2θ = 10.2° that is linked $\lambda$ Signifies the wavelength of the X-ray beam, $\beta$ shows the full-width half maximum (FWHM) and $\theta$ displays the Bragg's angle.

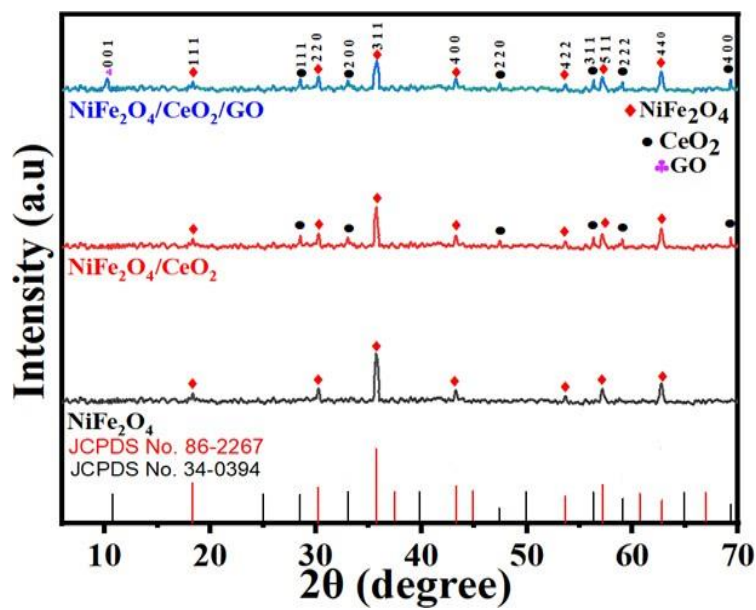

Figure 2: X-ray diffraction represents the crystalline structure of NiFe2O4, NiFe2O4/CeO2, and NiFe2O4/CeO2/GO nanocomposite to the interlayer space made possible by the oxygen functional group [61]. The Debye Scherrer's formula is used to determine the crystallite size of the prepared samples.

Figure 3 displays the morphological images from SEM of the prepared materials. In Fig. 3 (a), NiFe$_2$O$_4$ NPs were spherical, but the evident difference between nano-enabled NiFe$_2$O$_4$ and CeO$_2$ was visible. Moreover, NiFe$_2$O$_4$/CeO$_2$ exhibits some agglomeration processes as the metallic nanoparticles are easily agglomerated due to hydrophilic hydroxyl saturated residual bonds in different states on their surface as perceived in Figure 3 (b). Figure 3 (c) demonstrates the spherical nanoparticles of NiFe$_2$O$_4$/CeO$_2$ are arranged at the outermost layer wrinkles and GO interlayer structures. Additionally, NiFe$_2$O$_4$/CeO$_2$ has excellent dispersion and is well coated on the graphene oxide nanosheets, demonstrating the excellent performance of the hydrothermally synthesized NiFe$_2$O$_4$/CeO$_2$/GO nanocomposite [62]. The EDX spectrum of NiFe$_2$O$_4$, NiFe$_2$O$_4$/CeO$_2$, and NiFe$_2$O$_4$/CeO$_2$/GO nanocomposite are revealed in Figure 3 (d, e, & f). EDX analysis verified the high purity of the produced materials by confirming the presence of Ni, Ce, Fe, C, and O elements in the NiFe$_2$O$_4$/CeO$_2$ and NiFe$_2$O$_4$/CeO$_2$/GO nanocomposite and the FE-SEM and EDAX of CeO$_2$ in supplementary file Figure S2.

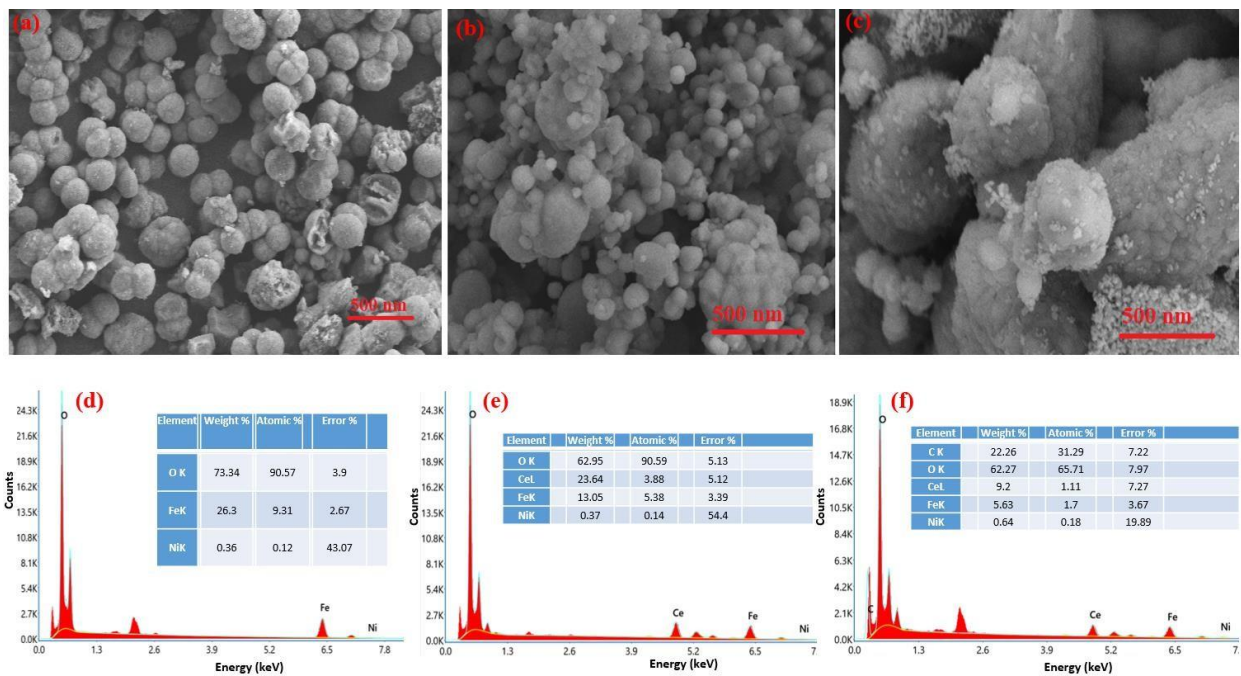

Figure 3: Morphological images of (a) NiFe2O4 (b) NiFe2O4/CeO2 (c) NiFe2O4/CeO2/GO exposed by FE-SEM and elemental composition revealed by EDAX of (d) NiFe2O4 (e)

NiFe$_2$O$_4$/CeO$_2$ (f) NiFe$_2$O$_4$/CeO$_2$/GO NC. Functional group analysis is represented by FTIR Spectroscopy as seen in Figure 4(a). The emergence of two major metal-oxygen bands between wavenumber 1000 cm$^{-1}$ and 400 cm$^{-1}$ provides evidence of ferrite's distinctive spinel structure. A prominent peak is shown in NiFe$_2$O$_4$ at 650 cm$^{-1}$ for the intrinsic tetrahedral metal-oxygen [Fe-O] vibrations (v1) and at 542 cm$^{-1}$ for the octahedral metal-oxygen [Ni-O] vibrations (v2). Based on the geometrical arrangement of the oxygen's closest neighbors, earlier reports indicate that the metal ions in ferrites are arranged in two distinct lattice sites, one tetrahedral and the other octahedral. The ferrite's vibrational spectra have a high frequency between 680-600 cm$^{-1}$, which resembles the inherent vibrational mode of tetrahedral, and a low frequency between 550-400 cm$^{-1}$, which corresponds to the octahedral site [63]. In the spectrum, (O-H) stretching vibration shows the high-frequency range at (3200-3600) cm$^{-1}$ and the less intense broadband at (1500-1550) cm$^{-1}$ for all prepared samples [64]. The band at 1315 cm$^{-1}$ causes C-O stretching vibrations for NiFe$_2$O$_4$ and NiFe$_2$O$_4$/CeO$_2$/GO, respectively [65]. It is generally known that for nanoparticles, the characteristic vibrational frequencies of a functional group will change if the surroundings of the group are even slightly changed. As grain size decreases, higher vibrational frequencies are observed [66]. In the NiFe$_2$O$_4$/CeO$_2$/GO spectra, the graphene oxide possesses functional groups like O-H, C-OH, COOH, and C-O.

The carboxyl O-H stretching mode is responsible for the GO's IR spectral peak between 3445cm$^{-1}$ and 2568cm$^{-1}$. The incorporation of alcohols with water molecules causes the O-H stretching absorption peak (3400 cm$^{-1}$) which is overlaid on the carboxylic acid's OH stretch to occur [67]. The asymmetric and symmetric CH$_2$ stretching of GO is responsible for peaks relating to 2968 cm$^{-1}$ and 2849 cm$^{-1}$, respectively [68]. Figure 4 (b) demonstrates the Raman analysis to identify structural flaws in carbon-based materials. In ferrites, spinel A and B-site ions contribute to the Raman modes such as Eg, A1g, 3T2g, A1g, Eg, and T2g are associated with octahedral and tetrahedral cations in NiFe$_2$O$_4$/CeO$_2$/GO nanocomposite [69]. Oxygen symmetries in tetrahedral AO4 groups are responsible for peaks above 600 cm$^{-1}$. In NiFe2O4, the peak at 690 cm$^{-1}$ is attributed to Ag symmetry, while the peak around 320 cm-1 signifies that Eg symmetry may be found in oxygen. Octahedral group vibrations are attributed to the frequencies of 470 and 570 cm-1 for the F2g mode. Ce-O (Fluorite structure, F2g) symmetric stretching mode was perceived at 458 cm$^{-1}$ and 460 cm$^{-1}$, while defect induced oxygen vacancies mode at 596 cm$^{-1}$, and second order transverse acoustic (2TA) mode at (220 and 270 cm$^{-1}$) with the O-O stretching vibration for NiFe$_2$O$_4$/CeO$_2$ [70].For NiFe$_2$O$_4$/CeO$_2$/GO, D and G peaks are identified from graphene at 1360 cm$^{-1}$ and 1595 cm$^{-1}$, respectively [71, 72]. However, Carbons with sp3 hybridization are presented by the D band, while sp2 hybridization is revealed by the G band.

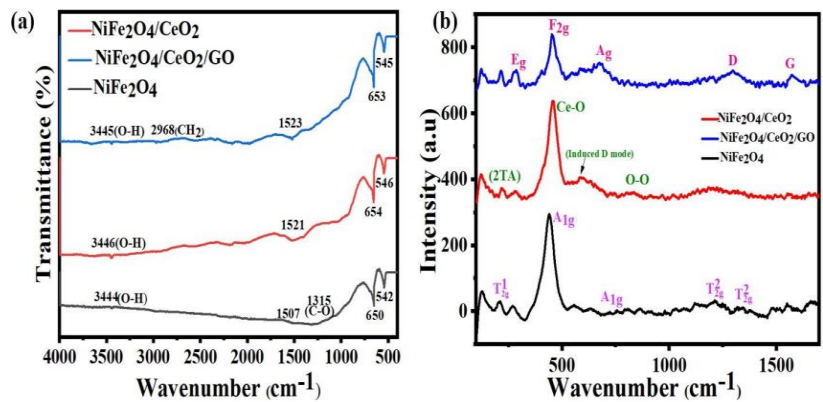

Figure 4: (a) FTIR Spectroscopy (b) Raman Spectroscopy of NiFe2O4, NiFe2O4/CeO2, NiFe2O4/CeO2/GO

In Figure 5 (a), Absorption spectra between 200 to 800 nm were assessed using UV-visible spectroscopy which represents the optical characteristics of the $NiFe_2O_4/CeO_2$, $NiFe_2O_4$, and $NiFe_2O_4/CeO_2/GO$ NC [73] and the absorbance graph of CeO2 in Supplementary file Figure S3. The occurrence of cations ($Fe^{3+}$ & $Ni^{2+}$) in the structure of the ferrite is indicated by a prominent absorbance peak in the visible area spectra between 250 and 700 nm [74]. It indicates that increased light absorption is useful to generate more electron-hole pairs, which may result in increased photocatalytic activity. Also, $NiFe_2O_4/CeO_2$ nanocomposite has a noteworthy absorption intensity in the visible range. In $NiFe_2O_4/CeO_2/GO$ nanocomposite, the absorption spectrum displayed a peak corresponding (n→ $\pi*$) transitions for (C=O) and ($\pi \rightarrow \pi*$) transitions for aromatic (C-C) bonds which may occur due to the deoxygenation of the graphene oxide [75]. The band gap of synthesized materials was evaluated by the following formula [76].

$$(\alpha h\nu)^n = A(h\nu - E_g)$$

where $\alpha$ is absorbance constant, $h\nu$ indicates the energy of an incident photon, $Eg$ represents band gap energy, Plank's constant (h), $k$ has constant values. Plotting $(\alpha h\nu)2$ vs $h\nu$ revealed the direct band gap of synthesized materials. Calculated 'Eg' values of CeO2 (3.2 eV), $NiFe_2O_4$ (1.95 eV), $NiFe_2O_4/CeO_2$ (2.7 eV), and $NiFe_2O_4/CeO_2/GO$ (2.68 eV) NC are shown in Figure 5 (b).

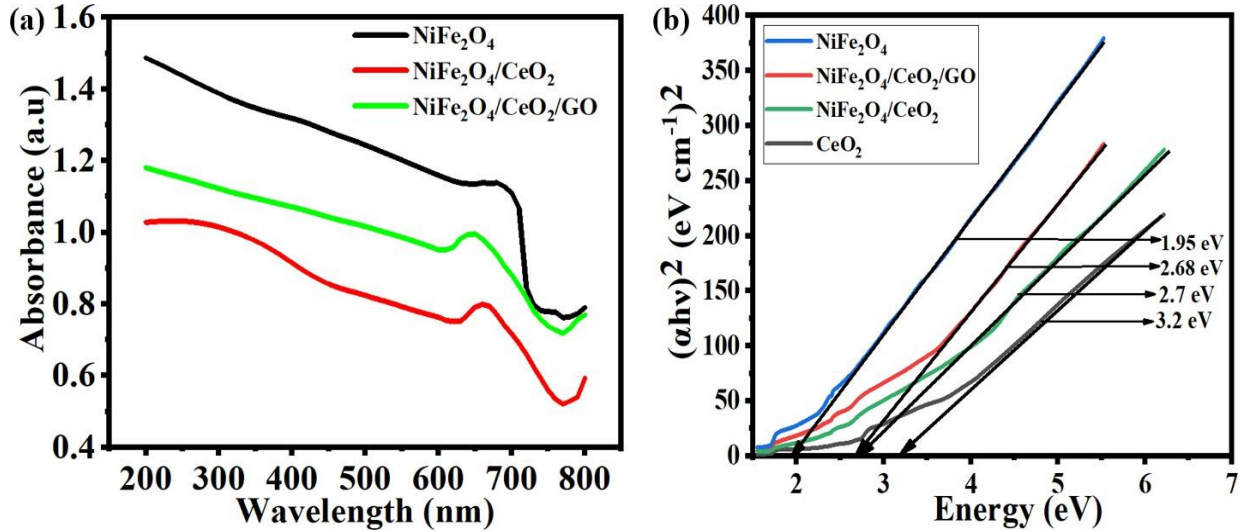

Figure 5: UV-visible Spectroscopy for NiFe2O4, NiFe2O4/CeO2, and NiFe2O4/CeO2/GO (a) Absorption spectra of (b) bandgaps calculated of all prepared materials via tauc plot.

In Figure 6 (a), VSM examined the magnetic hysteresis curves of the $NiFe_2O_4/CeO_2$ and $NiFe_2O_4/CeO_2/GO$ NC [77]. Shape, crystallinity, and the evolution of magnetization are all factors that contribute to the persistence of magnetic properties in magnetic materials [78]. At ambient temperature, the $NiFe_2O_4/CeO_2$ displays normal ferromagnetic activity, with a saturation magnetization (Ms) of around 34.79 emu/g [79]. $NiFe_2O_4/CeO_2/GO$ nanocomposite has a saturation magnetization (Ms) value of 12.21 emu/g. Since GO was added with $NiFe_2O_4/CeO_2$, therefore, it is expected that the magnetic component is substantially lower in GO, resulting in a lower total magnetization value. Coercivity and magnetization of $NiFe_2O_4/CeO_2$ are reduced when coated with GO, as the interaction between particles is reduced [80]. Magnetic parameters are calculated by the following equations as listed in Table S1.

The Brunauer Emmett Teller (BET) technique was used to estimate the specific surface area of each sample, and the Barret Joyner Halenda (BJH) method was used to calculate the mean diameter of the pores and total volume. The structural porosity of the samples was examined using these metrics. The nitrogen-gas adsorption /desorption isotherms' hysteresis loop is classified by the UPAC as illustrated in Figure 6 (b), and the associated inset BJH curves indicate the presence of mesoporous (2-25 nm) structure [81, 82]. Hysteresis loops, a feature of mesoporous structures, may be seen in the adsorption-desorption curves of all produced materials. The outcomes reveal that the specific surface area increases when GO is added, rising from 92 $m^2.g^{-1}$ to a maximum of 160 $m^2.g^{-1}$. The porous sheets of GO, a supporting layer, are responsible for the rise in the specific surface area. Nitrogen-gas adsorption /desorption isotherms' hysteresis loop of NiFe2O4, NiFe2O4/CeO2, and different surface area, pore volume, and pore sizes of all prepared samples are mentioned in supplementary Figure S4 and Table S2. To examine electrical conductivity and varied impedances, EIS was conducted between 0.1 Hz and 105 Hz in the frequency range. The comparable Nyquist curve of the electrode NiFe2O4, NiFe2O4/CeO2, and NiFe2O4/CeO2/GO EIS test is shown in Figure 6(c). The analogous series resistance (Rs) is displayed at the x-intercept. This resistance is made up of the junction resistance at the electrode material/substrate interface, the ionic resistance from the electrolyte, and the intrinsic resistance of the electrode material and substrate [83]. The lower Rs value of 1.1 Ω demonstrates highly appreciable electrical conductivity of the NiFe2O4/CeO2/GO electrode as compared to NiFe2O4, NiF2O4/CeO2. The small diameter of the semi-circle implies a lower charge transfer resistance (Rct) of 2.2 Ω. The high structural mesoporosity of NiFe2O4/CeO2/GO electrode materials shortens the charge transmission path across the electrode interface and KOH electrolyte solution [84,85].

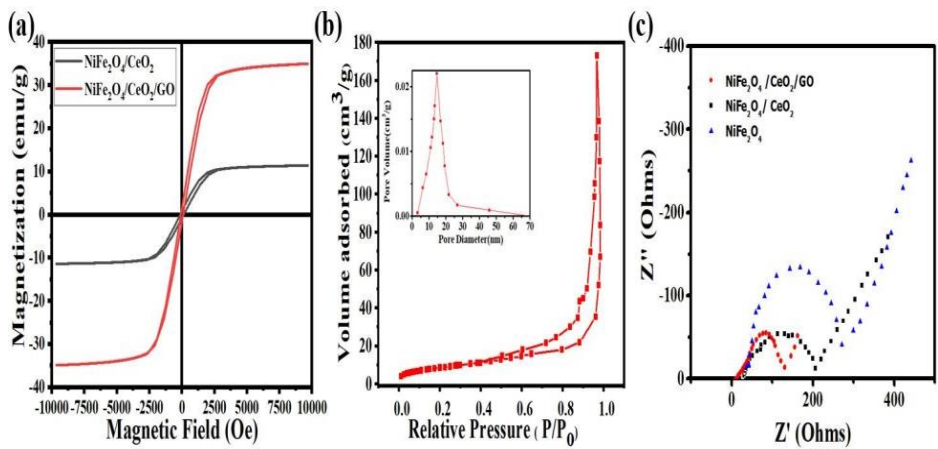

Figure 6: (a) Magnetization hysteresis behavior of NiFe2O4 and NiFe2O4/CeO2/GO NC (b) nitrogen-gas adsorption /desorption isotherms' hysteresis loop of NiFe2O4/CeO2/GO NC (c) EIS graph of NiFe2O4, NiFe2O4/CeO2 and NiFe2O4/CeO2/GO NC

Fig. 7a clarifies the survey spectrum of (Ni, Ce, Fe, C, and O), which together make up the NiFe2O4/CeO2/GO NC. XPS study explains the existence of differential elements and metallic species in NiFe2O4/CeO2/GO composite. The Ni 2p3/2 and Ni 2p1/2 XPS spectra are shown in Fig.7b. The primary peak of the Ni 2p3/2 spectrum is resolved into two distinct peaks located at 855.6 eV and 857.0 eV, respectively. These two peaks correspond to the valence states of Ni2+ and Ni3+, respectively. Furthermore, two satellite peaks are observed at energy levels of 863.3 eV and 880.3 eV. The locations of the deconvoluted and satellite peaks are in agreement with the values that have been reported [86]. In addition, the Ni 2p spectra exhibited two main peaks, namely Ni 2p1/2, with signals centered at 872.5 eV, and 874.6 eV, which are indicative of Ni3+ and Ni2+ features, respectively [87]. Fig. 7c displays the Ce spectra of five distinct peaks with binding energies of 890.9 eV and 904.4 eV, which correspond to the Ce 3d5/2 and Ce 3d3/2 states, respectively. These states indicate the valence state of Ce3+. Furthermore, the peaks observed at the binding energies of 882.5 eV can be attributed to Ce 3d5/2, whereas the peaks at 897.7 eV and 910.6 eV are associated with Ce 3d3/2, indicating Ce4+ oxidation states [88]. Fig. 7d displays the core level Fe 2p3/2 spectra, which reveal two prominent peaks located at 710 eV and 724.5 eV. These peaks correspond to the valence states of Fe2+ and Fe3+, respectively, thereby verifying the successful functionalization of Fe2O4. The presence of Fe2+ and Ni3+ can be attributed to electron transfer between Ni and Fe ions (Fe3+ + Ni2+ ↔ Fe3+ + Ni2+). Similar findings have been reported for Co- Ti-doped M-type strontium hexaferrites [89]. The sp2 hybrid C=C, sp3 hybrid C-C, and O-C=O of (C-1s) spectra peaks appear at (284.6, 287.47, and 288.5 eV) after deconvoluting the raw data to find overlapping peaks in Fig. 7e [90]. In Fig. 7f, numerous peaks in O-1s spectra that occur at (529, 531, and 532 eV) respectively, are attributed to oxygen, which exists in corresponding ester, hydroxyl, and carbonyl functional groups [91].

The primary peak at 530.7 eV represents the combination of lattice oxygen $O^{2-}$ with Ni, $Ce^{4+}$, and $Fe^{2+}$. On the other hand, the peak at 533.2 eV refers to the presence of adsorbed oxygen in the catalyst. The oxygen that is adsorbed can produce free radical active species by trapping electrons that are created by light during the photocatalytic process. This helps to enhance the activity of the photocatalyst [92]. Therefore, according to Fig. 8b, $NiFe_2O_4/CeO_2/GO$ nanocomposites exhibit superior adsorption performance compared to $NiFe_2O_4$ and $NiFe_2O_4/CeO_2$. This enhanced adsorption capability can contribute to the improvement of activity in the first stage of photocatalysis. Some of the $Ce^{4+}$ in $CeO_2$ may catch electrons produced by light to form $Ce^{3+}$. By incorporating $Ce^{3+}$ into the $NiFe_2O_4$ structure, a new impurity level is created within the band gap, resulting in a reduction in the width of the band gap. Conversely, a heterojunction is created when $CeO_2$ and $NiFe_2O_4$ are combined, resulting in an expanded range of spectrum response and enhanced photocatalytic activity of the catalyst. The XPS figure (b,c&d) demonstrates that the photocatalytic efficiency of $NiFe_2O_4/CeO_2$ was enhanced when compared to $CeO_2$ and $NiFe_2O_4$. The combination of $NiFe_2O_4$ with $CeO_2$ forms a heterojunction, which enhances the separation of photo-generated electron-hole pairs and thus enhances the performance of composite photocatalyst materials [93]. The presence of graphene oxide (GO), which has high conductivity, enhances the movement of carriers at the interface and enhances the photocatalytic activity of the composite catalyst. Hence, $NiFe_2O_4/CeO_2/GO$ exhibits superior adsorption and photocatalytic capabilities, resulting in a TC-HCl removal rate of 95 % after a 90-minute photoreaction.

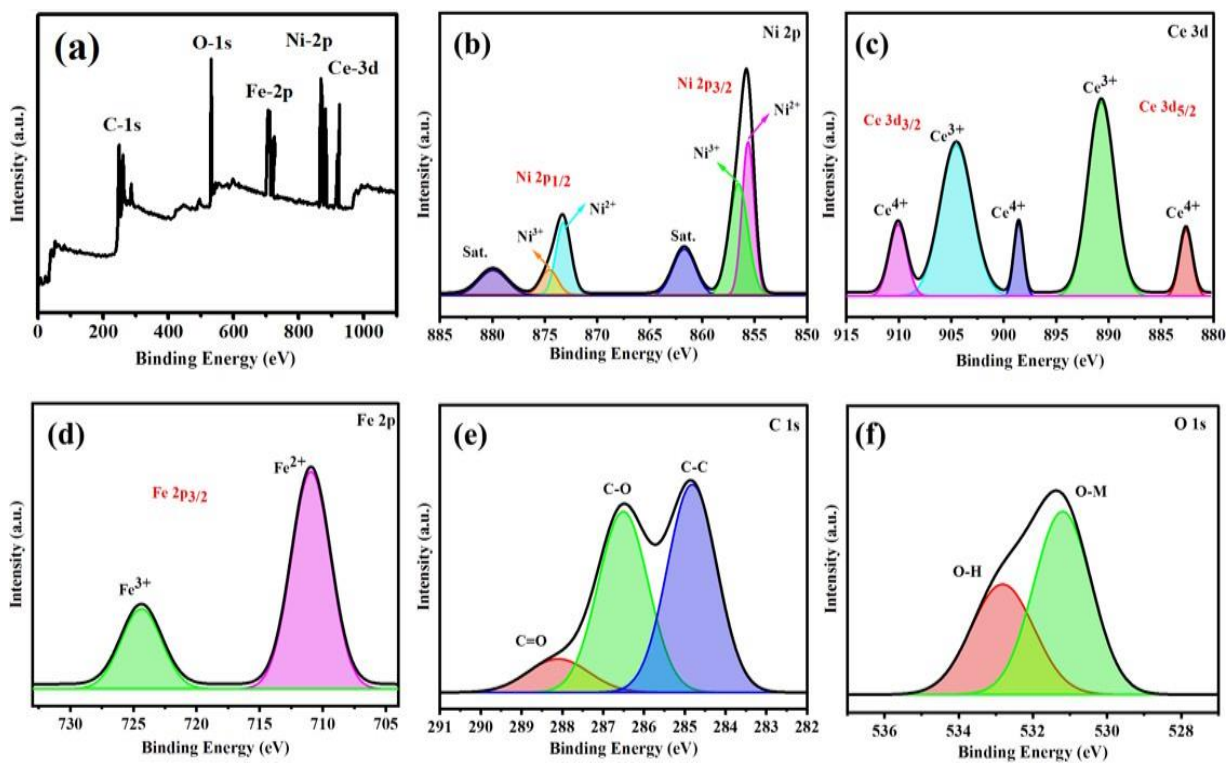

Figure 7: High-resolution XPS spectrum of NiFe2O4, NiFe2O4/CeO2 and NiFe2O4/CeO2/GO nanocomposite (a) complete spectrum (b) Ni 2p (c) Ce 3d (d) Fe 2p (e) O-1s (f) C-1s.

3.2 Photocatalytic performance, stability, and photodegradation of TC-HCl: A small amount of physical tetracycline-hydrochloride (TC-HCl) elimination by adsorption shows almost no effectiveness in the dark as shown in Figure 8 (a). Besides, following 90 minutes of visible light exposure, the degradation of TC-HCl in the photocatalyst's absence may also be ignored which indicates the photocatalysts and exposure to visible light are necessary for the degradation of tetracycline-hydrochloride. Alternatively, NiFe2O4/CeO2/GO nanocomposite enhances the TC-HCl photodegradation performance to 95% as shown in Figure 8 (b). It is obvious that the interaction of NiFe2O4/CeO2 with TC stimulates the production of numerous charge carriers when exposed to visible light and further by adding the GO, further suppresses electron- hole pair interaction. The improved photocatalytic efficacy of the nanocomposite catalysts has been related to a reduced rate of recombination between charge carriers and rapid transportation of photoinduced electron-hole couples. The following equation, which represents the pseudo-first-order kinetics model, is frequently used to demonstrate tetracycline degradation when exposed to photocatalysts.

$-\ln (C/C_0) = kt$

The rate constant (k) for photocatalytic degradation was determined using C and C0 as starting and final concentrations. In Figure 8 (c&d), rate constant (k) was calculated to be 0.015 min-1 for CeO2, 0.019 min-1 for NiFe2O4, 0.031 min-1 for NiFe2O4/CeO2, and 0.067 min-1 for the nanocomposite of NiFe2O4/CeO2/GO. The ability of nanocomposite materials to be recycled is essential in environmental remediation studies. The performance of the NiFe2O4/CeO2/GO nanocomposite to degrade tetracycline after five cycles of reusability is depicted in Figure 8 (e), and it shows that the catalyst retains its 95% photodegradation performance, demonstrating its superiority as a catalyst. The catalytic results provide evidence that photocatalysts may be employed in practical wastewater treatments [94].

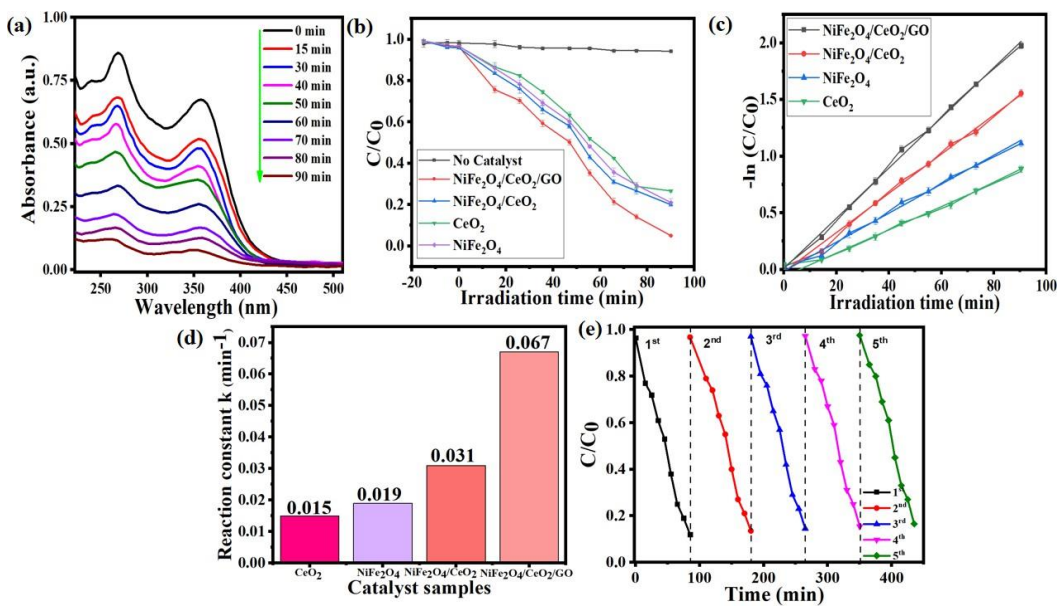

Figure 8: (a) variations in the tetracycline degradation show prominent absorbance at various times, (b) tetracycline Activity by using the prepared photocatalysts under visible light, (c) the kinetic plot using pseudo first order (d) all photocatalyst rate constant for TC-HCl (e) NiFe2O4/CeO2 /GO photocatalyst stability investigation for TC-HCl degradation.

To ascertain the optimum starting TC-HCl concentration, the effectiveness of photocatalytic degradation was seen in Figure 9 (a) at TC-HCl concentrations between 10-50 mg/L. Figure 9 (a) shows the photocatalytic breakdown of TC-HCl to be reduced from 95% to 83% with an increase in the starting TC-HCl dose from 10 mg/L to 30 mg/L. It appears that the catalyst's catalytic performance is also influenced by the TC's initial concentration. However, when the TC dose was increased from 30 mg/L to 50 mg/L, it became apparent that the reason for the effectiveness of the degradation might be that the higher initial concentration of TC caused the NiFe2O4/CeO2/GO active site to become saturated, resulting in less efficient degradation. [95]. During the TC-HCl photocatalytic process, pH may affect the stability of active species, the surface charge of catalysts, and the molecular structure of TC. The initial TC degradation efficiency was observed to decrease from 90% at pH 5 in Figure 9 (b). The outcomes showed that the maximum photocatalytic activity for TC breakdown was displayed by NiFe2O4/ CeO2 /GO at pH=5.

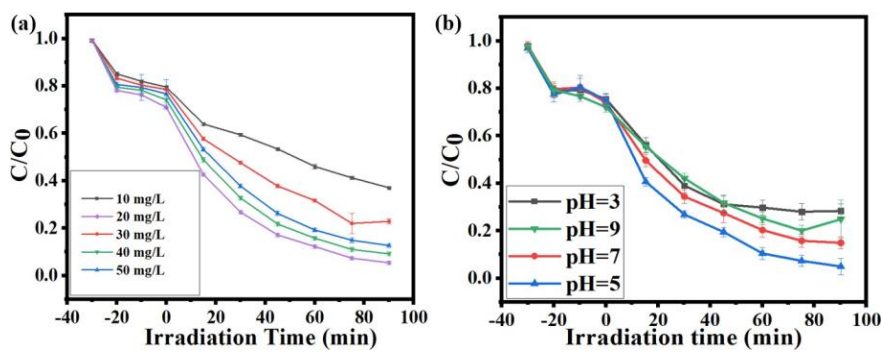

Figure 9: (a) TC-HCl curves analysis via different catalysts (b) photocatalysis Breakdown of TC-HCl for various pH ranges

Moreover, the impact of co-existing anions, such as $H_2PO_4^-$, $Cl^-$, $NO_3^-$ and $HCO_3^{2-}$, on the degradation of TCH-Cl as mentioned in Figure S5. The co-existing anions hindered the photocatalytic degradation efficiencies, and the negative impact on TCH-Cl degradation declined in the following order: $HCO_3^- > H_2PO_4^- > Cl^-, > NO_3^-$. The co-existing anions may function as free radical scavengers, decreasing the effects of dominant oxidizing radicals. Also, some anions may be adsorbed onto the NiFe2O4/CeO2/GO photocatalyst's surface, lowering the degradation efficiency [96, 97, 98]. Also, TCH-Cl degradation efficiency is still greater than 88% when coexisting inorganic cations (such as $Ca^{2+}$, $Mg^{2+}$, or $Na^+$) are present, as shown in Fig. S6. This suggests that the major metal cations do not inhibit the Photodegradation of TC-HCl via NiFe2O4/CeO2/GO nanocomposite. The reaction solution for TC-HCl degradation using the NiFe2O4/CeO2/GO nanocomposite was composed of three types of water matrices: pure water, tap water (Comsats University Islamabad, Lahore campus), and mineral water (Nestle) as seen in Figure S7. This was carried out to evaluate the impact of different water matrices for the practical usage of the photocatalyst in wastewater treatment. NiFe2O4/CeO2/GO catalyst degraded TC-HCl with efficiencies of 92%, 89%, and 90% for pure water, tap water, and mineral water, respectively [99].

### 3.3 Scavengers Radical trapping and electron spin resonance:

In Figure 10 (a), the radical trapping tests on scavengers allowed for the identification of the numerous photogenerated species during TC-HCl breakdown. Ammonium oxalate (AO) and silver nitrate (AgNO3) served as the scavengers for holes (h+) and electrons, while BQ and IPA performed as the scavengers for $•O^-$ and •OH radicals. In Figure 10 (b), nearly 90% of TC was removed when there were no scavengers. BQ substantially slowed down the TC-HCl's ability to degrade, showing that superoxide anion radicals were the most active species in the breakdown of tetracycline hydrochloride. However, the addition of IPA and AgNO3 showed a minimal effect on the TC-HCl photodegradation, indicating that hydroxyl radicals and electrons were also the significant reactive species for the tetracycline breakdown (degradation). In comparison, the presence of AO prevented the photocatalytic breakdown of TC-HCl, proving that holes were the primary active species [101]. The ESR procedure supports the emergence (generation) of photo-induced ($•O^-$ and •OH) radicals using DMPO serving as a trapping agent, as illustrated in Figure 10 (c & d) [102]. In both cases (DMPO $•O^-$ and DMPO •OH), ESR signals were evident within 5 and 10 minutes under the impact of visible irradiation, respectively while ESR signal was not noticed in the dark condition. Additionally, the signal strength increased by utilizing NiFe2O4/CeO2/GO and increasing the visible-light exposure to a duration of 10 minutes.

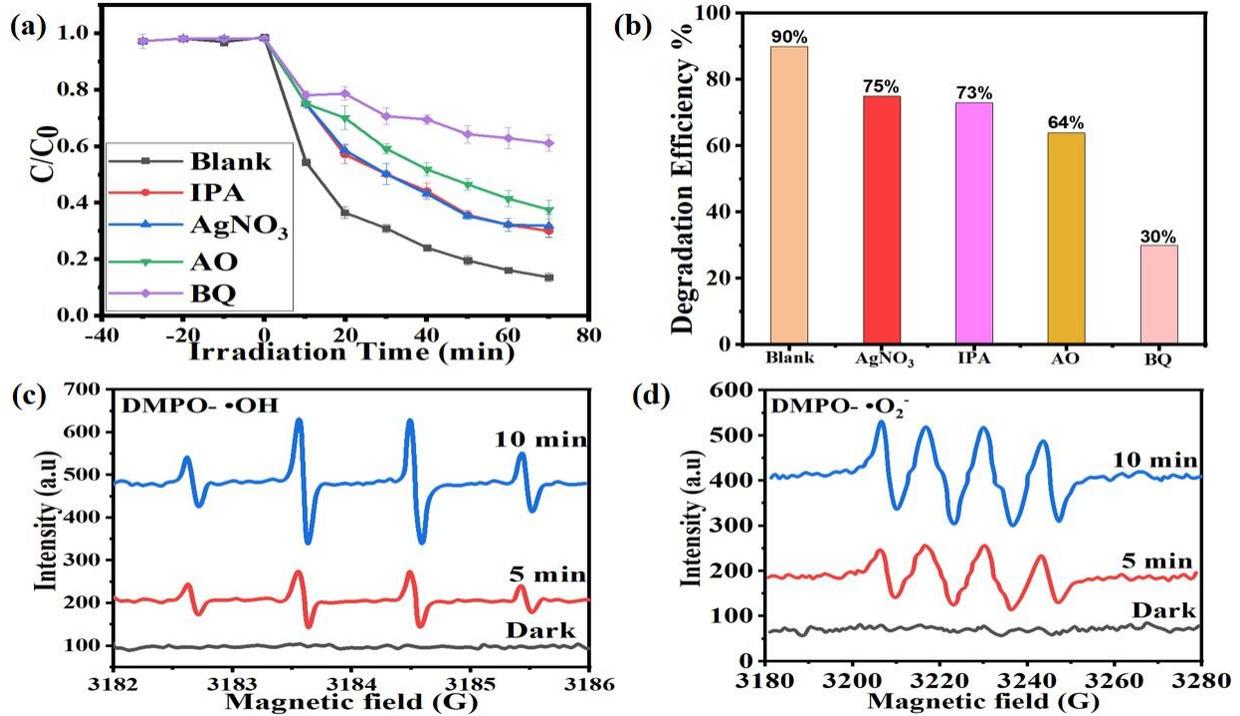

Figure 10: (a) Scavengers outcomes for TC-HCl breakdown degradation (b) Degradation efficiency of scavengers for tetracycline (c & d) ESR spectral signals for the hydroxyl and superoxide anions in both light and dark

Figure 11 displays the photocatalytic elimination of TC-HCl by the NiFe2O4/CeO2/GO nanocomposite. When NiFe2O4 and CeO2 nanoparticles are exposed to visible light, they cause the $e−/h+$ pair in the VB and CB, which is seen in the schematic diagram below in Figure 11. Therefore, the potentials of VB and CB were calculated by using the following equations [103].

$E_{CB} = X - (E_{ef} + 0.5 \, E_g)$ & $E_{VB} = E_{CB} + E_g$

Where the valence band and conduction band potentials, respectively, are denoted by $E_{CB}$ and $E_{VB}$. $E_{ef}$ (free electron energy) which is equivalent to 4.5 eV on the hydrogen scale. The following equation may be used to compute $E_g$, which stands for the semiconductor's band gap energy, and X, which stands for the semiconductor's electronegativity:

$X = [(A)^a x(B)^b x(C)^c]^{1/((a+b+c))}$

where a, b, and c represent the number of atoms in each compound [104]. The band gap energy ($E_g$) and X of NiFe2O4 and CeO2 are 1.95 eV, 5.83 eV, and 3.2 eV, 5.56 eV, respectively. As a result, NiFe2O4 and CeO2 had ECB and EVB values of 0.36 eV, 2.31 eV, and -0.6 eV, 2.7 eV, respectively, compared to the NHE (Normal Hydrogen Electrode). Additionally, Figure 11 offers a clear illustration of the impact of graphene oxide nanosheets, that trap photoinduced electrons when transit from the VB to the CB. Superoxide radical anion ($\bullet O^-$) is produced when molecular oxygen is reduced during the electron transfer. However, hydroxyl radicals ($\bullet OH$) are created when valence band holes join with water molecules or hydroxyl groups to trigger oxidation. By reacting with tetracycline, these reactive species ($\bullet OH$, $\bullet O^-$) remove the color from the compound. The photocatalytic removal of tetracycline is illustrated by the following equations:

$$NiFe2O4/CeO2 + h\nu \rightarrow NiFe2O4/CeO2(e^- + h^+)$$
$$NiFe2O4/CeO2(e^- + h^+) + GO \rightarrow NiFe2O4/CeO2(h^+) + GO(electron)$$
$$G(electron) + O2 \rightarrow \bullet O^- + GO$$
$$NiFe2O4/CeO2(h^+) + H2O \rightarrow NiFe2O4/CeO2 + H^+ + \bullet OH$$
$$NiFe2O4/CeO2(h^+) + OH^- \rightarrow NiFe2O4/CeO2 + \bullet OH$$
$$TC + (h^+, \bullet OH, \bullet O^-, e^-) \rightarrow intermediates \rightarrow CO2 + H2O (degradation\ products)$$

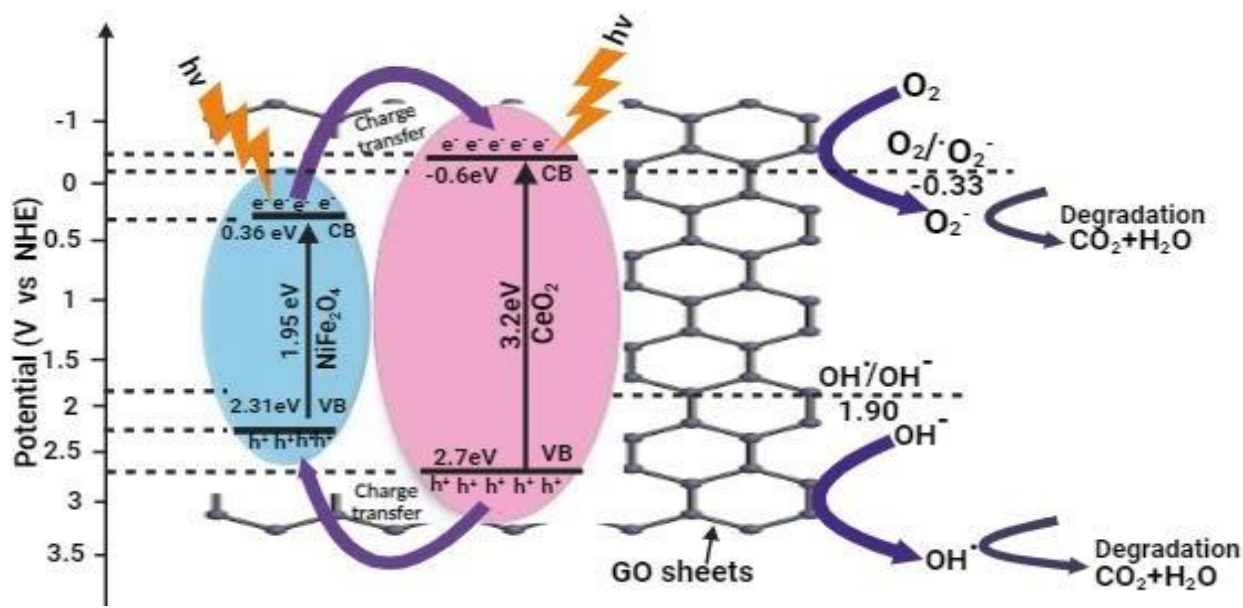

Figure 11: Schematic representation depicted the enhanced production of holes and superoxide anions during the photodegradation of tetracycline antibiotic using NiFe2O4/CeO2/GO

The stability of the material in practical applications is also a significant aspect that must be considered into account when selecting the most effective photocatalyst [105]. Thus, the stability of the NiFe2O4/CeO2/GO nanocomposite was assessed by the use of XRD and SEM measurements. As a result, the XRD pattern and SEM illustrations, which compare the phases and microstructure with those of the fresh one (Figure 12(a-c)), reveal the great stability of NiFe2O4/CeO2/GO. The aforementioned results indicate that NiFe2O4/CeO2/GO is a practicable material for wastewater treatment due to its large mineralization ability, superior stability, reusability, and high activity.

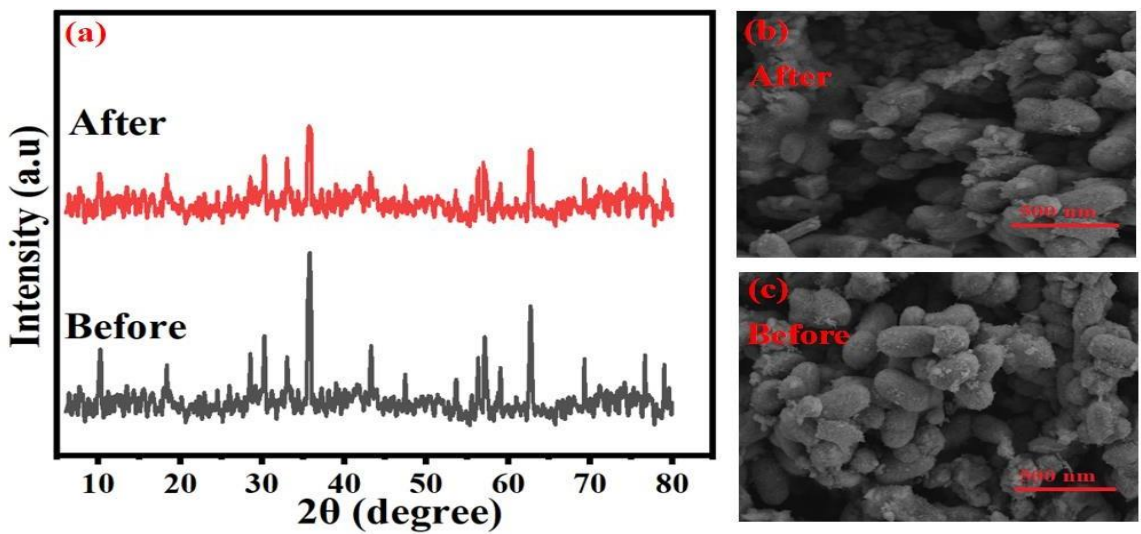

Figure 12: Stability analysis via (a) XRD and (b&c) SEM

**3.4: Possible degradation pathways, intermediates, and toxicity assessment test:**

Using an LC-MS system, the intermediates generated during the photocatalytic reaction were identified to get a better understanding of the NiFe2O4/CeO2/GO nanocomposite-induced TC degradation process. Past studies have shown that TC decomposes readily due to the easy assault of active free radicals on electron-donating groups and ionizable groups (such as dimethylamino groups, amide groups, and double bonds) [106,107]. Three primary TC degrading mechanisms over NiFe2O4/CeO2/GO are suggested in this work and shown in Figure 13. Since the C = C double bond in route 1 is very unstable and readily oxidized by •OH, the hydroxylation process results in P1 (m/z = 461) [108,109]. P4 (m/z =338) is then produced as a result of the deamination and ring-553   opening reactions, while P7 (m/z =307) is produced as a result of the subsequent de-hydroxylation process when h+, $O-$, and •OH are present [110]. Route 2: TC at m/z = 445 with $SO•-$ attack releases the –N(CH3)2 group, generating the P2 (m/z = 415) intermediate. Route 2 is another possibility. In this pathway, TC is subjected to dehydration, demethylation, and amino group loss, resulting in P5 (m/z = 318) and P8 (m/z = 246) [111,112]. Route 3 involves the easy assault of the amino group by h+ and •$O-$, which results in the creation of P3 (m/z =429) through the deamination reaction and subsequent degradation into P6 (m/z =399) through the demethylation process. Subsequently, a persistent oxidation process involving • $O-$ and •OH leads to a second degradation event that produces P9 (m/z = 302), in agreement with the earlier finding [113]. Finally, these intermediates are broken down and mineralized into tiny molecules by reactive species.

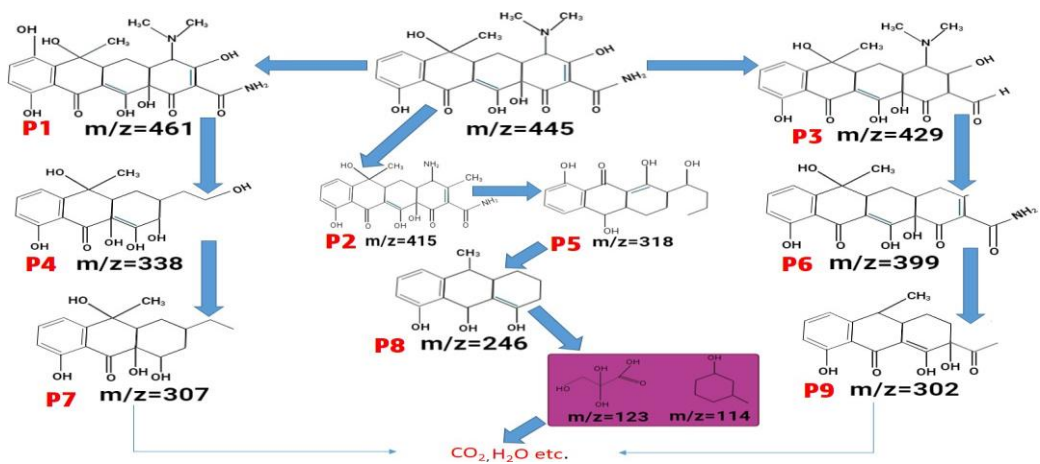

Figure 13: NiFe2O4/CeO2/GO nanocomposite revealed the LC-MS pathway for the TC-HCl breakdown into smaller intermediate molecules

According to QSAR prediction, Toxicity Estimation Software (T.E.S.T.) was used to further assess the toxicity of TC and above intermediates that were converted from TC [114, 115]. Based on the findings, the majority of the intermediates had decreased in toxicity following the catalytic process. The fathead minnow's LC50 for TC was 1 mg L-1, as seen in Figure 14 (a), which is considered toxic. Even so, the LC50 of fathead minnow for P-461, P-429, P-399, and P-338 was quite toxic, but the toxicity decreased to not be detrimental for the subsequent oxidation intermediates. The toxicity of TC for Daphnia magna rapidly declined, as seen in figure 14 (b) by the 48-hour LC50 Daphnia magna of 10 mg L-1, which was lower than that of nearly all the breakdown products. After development toxicity testing, it was found that half of the intermediates were less toxic to development than TC. In contrast, the other half was marginally more toxic to development than TC (Figure 14 (c)). According to the toxicity estimates above, the overall degradation process was low toxicity. It could deteriorate by raising the mineralization level and extending the reaction time, even while a tiny percentage of the intermediaries remained toxic. As a result, the combination demonstrates the ability of the NiFe2O4/CeO2/GO nanocomposite oxidation system to break down and lessen TC toxicity efficiently.

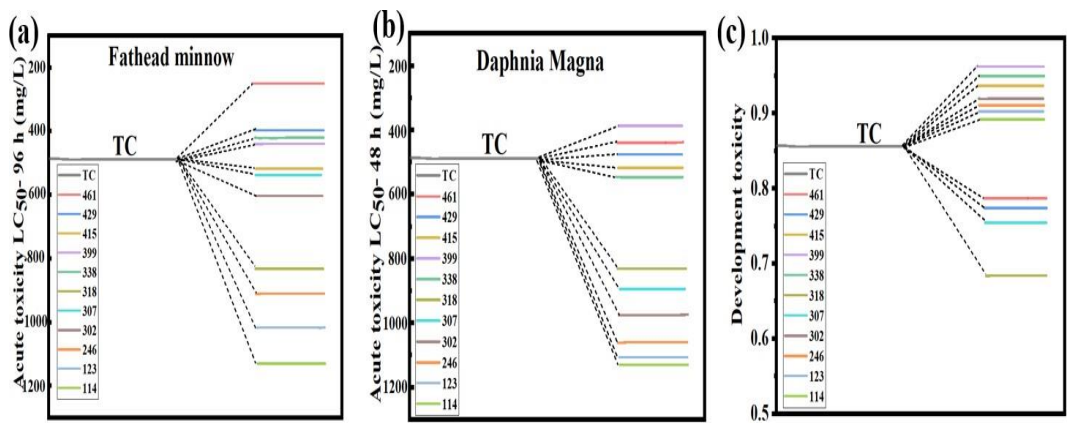

Figure 14: Acute toxicity for (a) Fathead minnow 96-h (b) Daphnia Magna 48-h (c) Development of toxicity of tetracycline and intermediates

Table 1: Comparison of the photocatalytic performance under visible light irradiation using different photocatalyst

| Photocatalyst | Types of Antibiotics/Pollutants | Amount of catalyst/Dosage (mg) | Volume (mL), concentration (mg/L) | Efficiency and time | Light Source | Ref. |
|---|---|---|---|---|---|---|
| $CoFe_2O_4@TiO_2@rGO$ | TC and CIP | 30 | 20,10 | 92% TC; 84% CIP in 90 min | Visible light (500 W Halogen lamp) | [116] |
| $\gamma$-$Fe_2O_3$/g-$C_3N_4$ | TC | 50 | 10 | 73.80% in 120 min | 500 W Xe lamp($\lambda >$ 420 $nm$) | [117] |
| ZnO/NiO/g-$C_3N_4$ | TC | 50 | 20 | 91.49% in 60 min | Visible light (500 W Halogen lamp) | [118] |
| $Fe_3O_4$/$BiVO_4$/CdS | TC | 100 | 10 | 87.37% in 90 min | 300 W Xe lamp | [119] |
| ZnO-$BiWO_6$/$Ti_3C_2$ MXene | CIP | 50 | 5 | 77% in 160 min | Natural light | [120] |
| $Ag_2CO_3$/$Bi_4O_5I_2$/g-$C_3N_4$ | TC | 30 | 20 | 82.61% in 30 min | 300 W Xe lamp | [121] |
| $NiFe_2O_4$/$CeO_2$/GO | TC | 50 | 20 | 95% in 90 min | 300 W Xe lamp($\lambda =$ 420 $nm$) | This work |

## 4. Computational Method

The Vienna Ab initio Simulation Package (VASP) implemented density functional theory (DFT) to calculate the structural and electrical characteristics of CeO2, NiFe2O4, and NiFe2O4/CeO2 [122, 123]. The wave functions are enlarged in plane waves with an energy cutoff of 550 eV using the projector augmented wave technique [124] to characterize the interactions between the valence and core electrons. When there is an energy difference of 10−6 or less between two successive iterations, the calculations were optimized using the conjugate gradient approach for geometric optimization. The Davidson method is used to minimize the Hellman-Feynman force, which is meant to converge if there is a difference of 0.01 eV/Ang between two successive iterations. This process achieves geometric relaxation. Reducible gamma-centered K-POINTS 10x10x10 are utilized for the bulk computation. The optimized kinetic energy cutoff, k-point, and lattice parameter optimizations were performed using the Perdew–Burke–Ernzerhof (PBE)computations to get precise structural and electrical characteristics [125].

### 4.1 Structural Optimization

Our computational investigation commenced with the bulk CeO2 structure, known for its fluorite structure characterized by the Fm3m space group. This configuration features a simple cubic oxygen sublattice, with Ce ions strategically positioned at alternating cube centers. The relaxed state of the structure is depicted in Figure 15 ((A) (a)), where the cerium ion is situated at the core of the tetrahedron, and oxygen atoms (depicted as red spheres) form the vertices. Subsequently, we proceeded to analyze the crystal structure parameters of the compound NiFe2O4 (Figure 15 (A) (b)), a spinel ferrite, using the volume optimization technique in the subsequent phase of our investigation. Typically, spinel ferrites are composed of a unit cell that possesses a face-centered cubic (fcc) lattice structure, where the cations are meticulously arranged at sites known as tetrahedral (Th) and octahedral (Oh) positions, referred to as A and B respectively, and are intricately coordinated with oxygen atoms. They are categorized as either normal- or inverse-phase based on how $X^{2+}$ (where X represents Ni, Mg, Co, Zn, Mn) and $Fe^{3+}$ are distributed between the Th and Oh sites. In the normal spinel ferrite structure, the Th sites are occupied by $X^{2+}$ cations, while the Oh sites are populated by $Fe^{3+}$ cations. The inverse phase emerges when the $Fe^{3+}$ cations fully occupy the Oh sites, and the Th sites are haphazardly filled with a combination of $Fe^{3+}$ and $Ni^{2+}$ ions. The extensively researched inverse spinel, nickel ferrite having a space group designation of 227, exhibits a distinctive arrangement. In this structure, Ni atoms are positioned at the 8a tetrahedral sites (1/8, 1/8, 1/8), Fe atoms occupy the 16d octahedral sites (1/2, 1/2, 1/2), and O atoms reside at the 32e face-centered cubic structure sites (u, u, u). Structural optimization has been done for NiFe2O4, and as a result, we obtained an optimized value of a0=8.365 Ang and u=0.27 employing the PBE method, which is in good agreement with the experimental data. Graphitic Oxide (GO) exhibits a structurally unique and less ordered configuration compared to pristine graphite. The absence of a well-defined space group underscores the amorphous nature of graphite oxide, arising from the introduction of oxygen-containing functional groups during the oxidation process. The construction of a 4x4x1 supercell of GO is undertaken to envelop the composite's surface area, yielding a lattice mismatch of 1.9%. The interlayer spacing between composite and GO was determined to be 3.06 Ang, aligning closely with reported values for GO-based photocatalysts [126, 127]. In the end, we built NiFe2O4/CeO2/GO composite design and structure optimization. The composite structure is designed by using Material Studio and optimized in VASP until reached required accuracy.

## 4.2 Electronic Properties

CeO2 exhibits the properties of a wide bandgap semiconductor, possessing a primary energy gap of ~2.42-3.0 eV. The conduction band is shaped by the Ce4f states, whereas the valence band is constituted by the O2p states. The effective mass of the electrons (me) is remarkably high as a result of the minimal dispersion exhibited by the f-bands. In contrast to charge transport in conventional semiconductors, where charge carriers can move within their respective bands, the electrons within CeO2 are highly localized on a Ce atom, giving rise to a small polaron. The density of states (DOS) assists in comprehending how states are distributed within specific energy ranges, offering insights into both unoccupied and occupied states. The states that can be occupied are described by a significant density of states at a particular energy level, but there are no states occupied when the density of states is zero. The density of states for CeO2 exhibits a discontinuity within the range spanning from the upper limit of the valence band to the lower limit of the conduction band, which typically corresponds to the system's bandgap. The band structures and density of states of CeO2 computed with the PBE approximation are illustrated in Figure 15 ((B) (a) and (b)), respectively. The density of states shown in Figure 15 ((B) (b)) arises from the incorporation of O2p orbitals (ranging from 5 eV to 9 eV), Ce4f orbitals (ranging from 0.5 eV to 1.7 eV), and Ce5d orbitals (ranging from -1.8 eV to -5 eV). The calculated bandgap of CeO2 spans from O2p to Ce4f with a width of 3.6 eV, and from O2p to Ce5d with a width of 6.8 eV. The acquired band structure, illustrated in Figure 15 ((C) (a)), unveils the lack of band intersections in the vicinity of the Fermi level. This observation allows us to infer that the calculated band structure of NiFe2O4 exhibits semiconductor characteristics, featuring a gap energy of approximately 1.67 eV, which aligns with the results obtained from experimental studies. Figure 15 ((C) (a-b)) shows the total density of states (TDOS) and partial density of states (PDOS) of individual atoms in NiFe2O4. The $O^{2-}$ states dominate the valence band, where they also set the band edge. The $Ni^{2+}$ states influence the VB to some extent, below the VB edge, and is determined by an overlap of $O^{2-}$ and $Ni^{2+}$ states. Except for a little contribution from the $O^{2-}$ states, the conduction band edge primarily depends on the $Fe^{3+}$ tetrahedral states. $Fe^{3+}$ octahedral states again dominate the CB, with little contribution from $O^{2-}$ states. Thus, these are the $Fe^{3+}$ tetrahedral and octahedral states that control the CB band edge in the inverse spinel NiFe2O4. In the end, the electronic structures of the isolated bulk structure of NiFe2O4/CeO2/GO were calculated, including the band structure and partial density of states (PDOS). An analysis of the electronic band structures of the composite revealed that its conduction band minimum (CBM) and valence band maximum (VBM) are positioned at the same high-symmetry Γ-S locations in the Brillouin zone, making it phases direct bandgap semiconductors. It can be observed from the predicted PDOS that the VBM consists mostly of Ce-5d and Ni-3p atomic orbitals, whereas the CBM principally of Fe-5s, O-3p, and C-2p orbitals. Meanwhile, a considerable hybridization between the Ce-4f state and the GO sheets' C-2p and O-2p states occurred at the CBM. The calculated band gap energy at 1.17 eV in (Figure 15 (D) (a-b)), the correctness and reliability of the theoretical approach and parameters used in this work were confirmed by the nanocomposites' consistency with the experimental results.

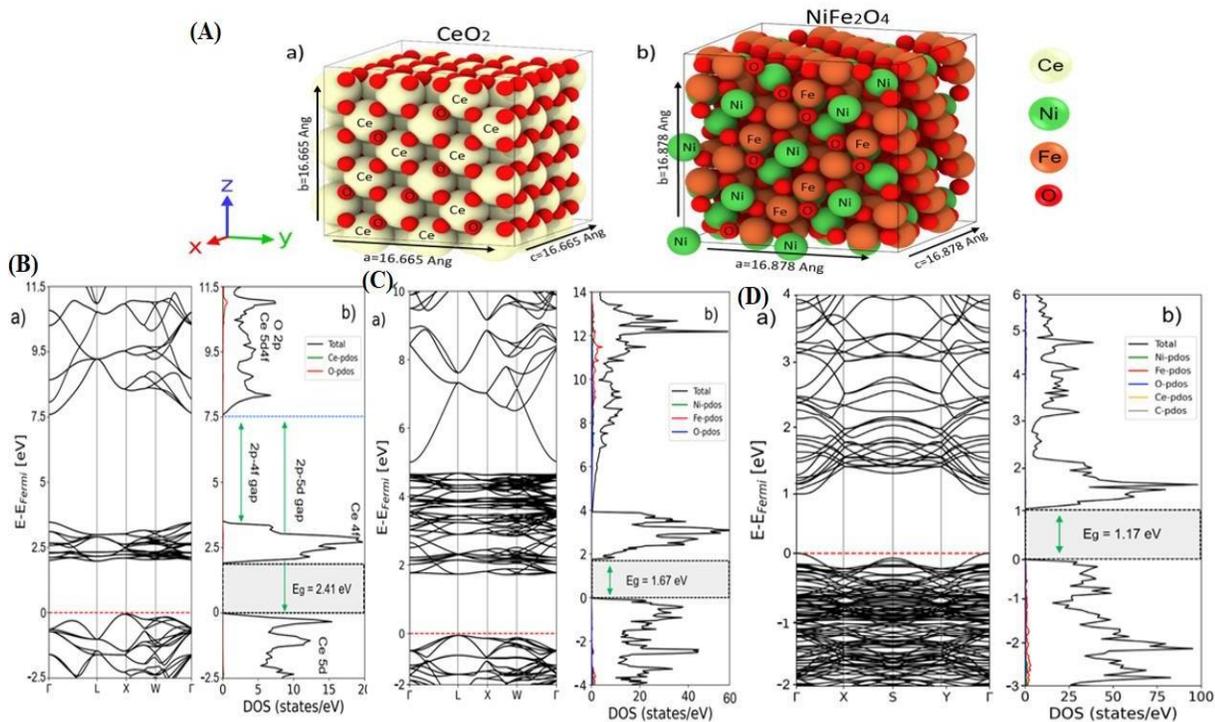

Figure 15 (A): The optimized structure of a) 3x3x3 CeO2 and b) 2x2x2 NiFe2O4 crystals. (B): Band structure and (partial) density of states of CeO2. The red line in the band structure plots represents the Fermi energy (C): Band structure and (partial) density of states of NiFe2O4 (D): Band structure and (partial) density of states of NiFe2O4/CeO2/GO.

## 5. Antibacterial activity

The antibacterial activity of NiFe2O4/CeO2/GO NC was evaluated with varying concentrations including 10 µg/mL, 20 µg/mL, 30 µg/mL, and 40 µg/mL against S. aureus and E. coli which are gram-positive and gram-negative bacterial strains respectively. The antibacterial activity was evaluated using the disc-diffusion method. As demonstrated in Figures 16(a) and 16(c) for S. aureus and E. coli, respectively, chloramphenicol was used as the positive control as a reference. It is seen that Figures 16(b) and 16(d) for S. aureus and E. coli show the antibacterial activity of NiFe2O4/CeO2/GO at concentrations of 10 µg/mL, 20 µg/mL, 30 µg/mL, and 40 µg/mL, which inhibited the inhibition zones. Furthermore, the antibacterial activity of NiFe2O4/CeO2/GO NC at different concentrations is compared using a bar graph that illustrates the dose-dependent behavior in Figure 16(e). The amount of inhibition for both bacterial strains increased with the concentration

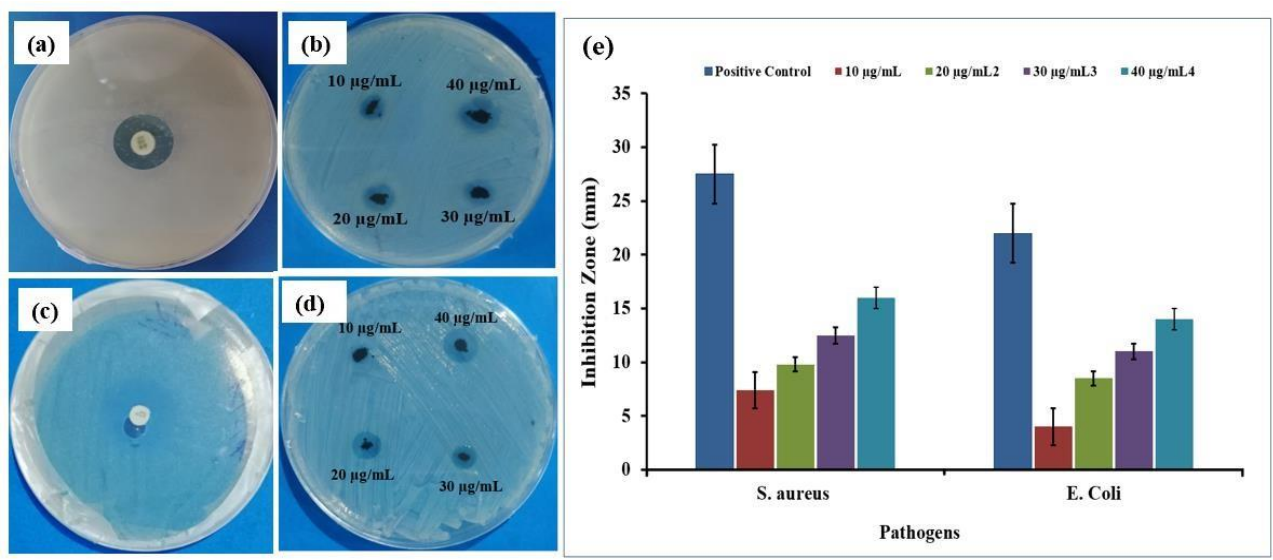

Figure 16: Antibacterial activity of (a) chloramphenicol against S. aureus strains (b) NiFe2O4/CeO2/GO with varying concentrations (10-40) µg/mL against S. aureus (c) chloramphenicol against E. coli (d) NiFe2O4/CeO2/GO with varying concentrations against E. coli (d) bar graph for comparison of NiFe2O4/CeO2/GO (10-40) µg/mL against S. aureus and E. coli

NiFe2O4/CeO2/GO nanocomposite showed a greater inhibition zone against S. aureus as compared to E. coli. NiFe2O4/CeO2/GO nanocomposite with a concentration of 40 µg/mL showed an inhibition zone of 16 mm and 14 mm against S. aureus and E. coli respectively. The greater inhibition of nanocomposite against S. aureus compared to E. coli is due to a complex phenomenon with multiple contributing factors such as Gram-positive bacteria have a simpler cell wall structure, primarily composed of peptidoglycan. The negatively charged peptidoglycan of S. aureus attracts positively charged nanoparticles through electrostatic interactions, facilitating adhesion and penetration. Herewith, NiFe2O4/CeO2/GO NC may damage the bacterial cell membrane through various mechanisms like direct contact, ion release, and reactive oxygen species (ROS) generation. ROS disrupts cell integrity and leads to leakage of essential cellular components, ultimately killing the bacteria.

### 5.1 Molecular Docking Analysis:

The application of computational tools to investigate the underlying mechanisms of biological activities and functions is widely reported [128,129]. Enzymes are widely recognized as the primary contributors to bacterial infection, and hence, suppressing their activity is crucial in combating the resulting infection [129]. The molecular docking analysis of the NiFe2O4/CeO2/GO nanocomposite against the FabI and DNA gyrase enzymes of E. coli and S. aureus demonstrated that they effectively block these specific enzyme targets. This was indicated by their strong binding affinity within the corresponding binding pockets. The docking analysis of

NiFe2O4/CeO2/GO NC with FabIE. coli demonstrated hydrogen bonding interactions with Thr194, Ile200, Ala197, Tyr156, Lys201, and Tyr146, resulting in a binding score of -7.2 kcal mol-1 (Fig. 17 (A) (a-c)). The NiFe2O4/CeO2/GO NC compound docked against FabIS. aureus showed a binding energy of -6.0 kcal mol-1. The interaction between the complex and the amino acids His247, Asn245, Lys172, Glu168, Ala165, Ala169, Glu151, and Lys172 was observed through hydrogen bonding, as depicted in Fig. 17 (B) (a-c). In addition, the binding potential of NiFe2O4/CeO2/GO NC was evaluated by analyzing the docking predictions for DNA gyrase from E. coli. The binding score was found to be -6.32 kcal mol-1. The evaluation focused on the active sites Ile78, Asn46, Ala47, Thr165, Val71, Asp73, Gly77, Val167, and Val43, as shown in Fig. 17 (C) (a-c).

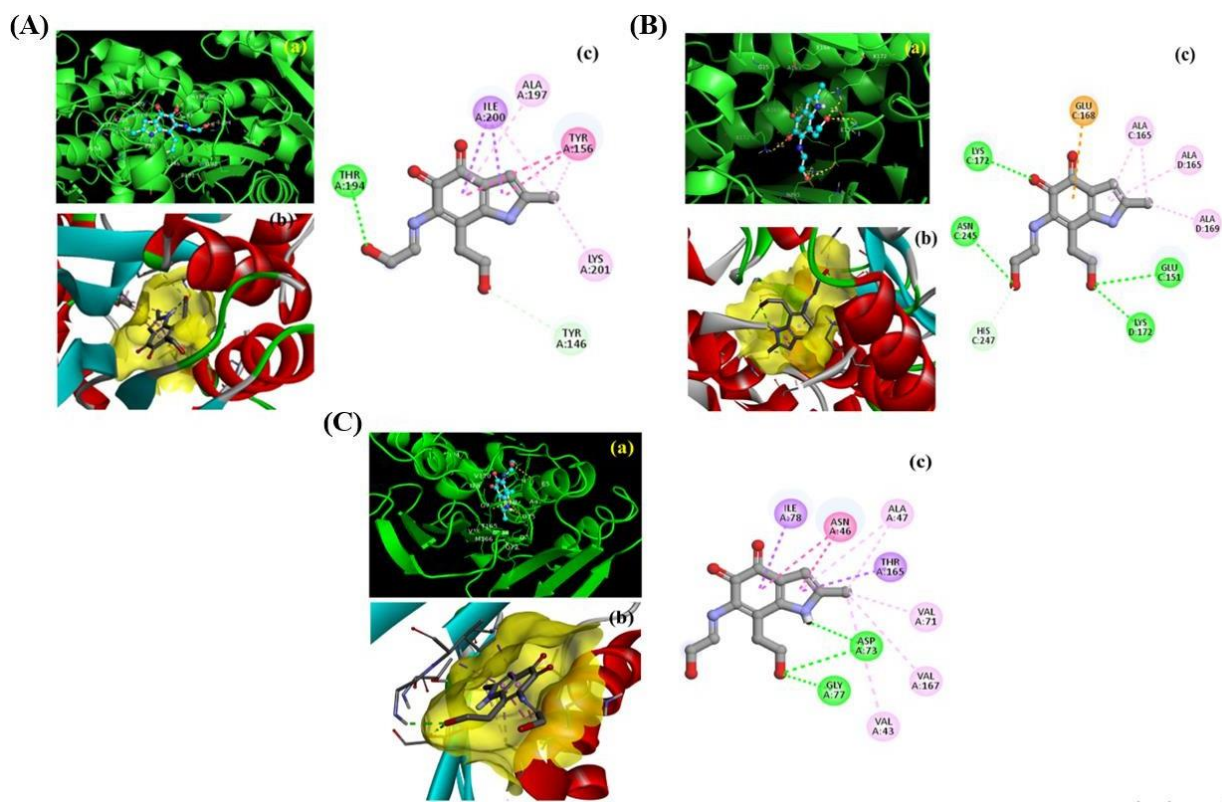

Figure 17: Binding interactions of NiFe2O4/CeO2/GO NC (a-b) (3D), (c) (2D)), inside the active pocket (a) FabIE.coli complex (b) FabIS. aureus complex (c) DNA gyrase from E. coli complex

**Conclusion**

In this study, NiFe2O4/CeO2/GO nanocomposite photocatalyst was prepared using a hydrothermal approach. The bactericidal activity of the NiFe2O4/CeO2/GO nanocomposite attributed to the inhibition of FabI and DNA gyrase, as revealed through in silico docking investigations. According to the current findings, NiFe2O4 and CeO2 were deposited on GO, which lowered the amount of photoinduced recombination of charged carriers, promoted charge carrier transport, and enhanced photocatalytic activity. NiFe2O4/CeO2/GO photocatalyst revealed an excellent TC removal with an efficiency of 95% in 90 minutes. Additionally, it was exposed that holes h+ and hydroxyl radicals prevailed in the breakdown of TC-HCl, and the NiFe2O4/CeO2/GO photocatalyst persisted in working well even after five cycles. After investigating the degradation intermediates of TC- HCl via LC-MS, the potential pathway in lower molecular intermediates and successive decomposition showed further to create carbon dioxide and water. The findings from the DFT investigation revealed that the contact interfaces between NiFe2O4 and CeO2 in the solid-solid state were established through chemical bonding rather than physical absorption. Adjustments in the CB and VB positions of NiFe2O4 and CeO2 occurred as a result of variations in the Fermi level, leading to the creation of a favorable structure. Overall, the NiFe2O4/CeO2/GO may serve as an excellent photocatalytic candidate nanocomposite for pharmaceutical degradation and may be practiced on a large scale for industrial purposes for water purification and environmental remediation.

**Acknowledgment**:
This work was supported by the Anhui provincial key research and development plan (Grant no. 202004i07020014). The authors would like to extend their sincere appreciation to the Researcher supporting program at King Saud University, Riyadh, for funding this work under project number (RSP2024R113).